# Time-resolved investigation of magnetization dynamics of arrays of non-ellipsoidal nanomagnets with a non-uniform ground state

P. S. Keatley, V. V. Kruglyak, A. Neudert, E. A. Galaktionov, and R. J. Hicken

*School of Physics, University of Exeter, Stocker Road, Exeter, EX4 4QL, United Kingdom*

J. R. Childress and J. A. Katine

*Hitachi Global Storage Technologies, San Jose Research Center, 3403 Yerba Buena Road, San Jose, California 95135, USA*

We have performed time-resolved scanning Kerr microscopy (TRSKM) measurements upon arrays of square ferromagnetic nano-elements of different size and for a range of bias fields. The experimental results were compared to micromagnetic simulations of model arrays in order to understand the non-uniform precessional dynamics within the elements. In the experimental spectra two branches of excited modes were observed to co-exist above a particular bias field. Below the so-called crossover field, the higher frequency branch was observed to vanish. Micromagnetic simulations and Fourier imaging revealed that modes from the higher frequency branch had large amplitude at the center of the element where the effective field was parallel to the bias field and the static magnetization. Modes from the lower frequency branch had large amplitude near the edges of the element perpendicular to the bias field. The simulations revealed significant canting of the static magnetization and the effective field away from the direction of the bias field in the edge regions. For the smallest element sizes and/or at low bias field values the effective field was found to become anti-parallel to the static magnetization. The simulations revealed that the majority of the modes were de-localized with finite amplitude throughout the element, while the spatial character of a mode was found to be correlated with the spatial variation of the total effective field and the static magnetization state. The simulations also revealed that the frequencies of the edge modes are strongly affected by the spatial distribution of the static magnetization state both within an element and within its nearest neighbors.

# 1. INTRODUCTION

During the past decade the magnetization dynamics of novel thin film magnetic materials have been studied intensely since these processes are expected to underlie the high frequency operation of future recording head sensors[1], magnetic random access memory elements[2], and spin-torque devices[3]. Experiments have been performed in which magnetization precession has been induced by harmonic[4] or pulsed[5] magnetic fields, ultrashort optical pulses[6], and a dc spin polarized current[7]. In the majority of such studies, the samples were essentially unbounded and the ground state magnetization was virtually uniform. Consequently, the precession frequency was fully determined by the magnetic parameters of the samples and the value of the applied (bias) magnetic field[8]. In contrast, the internal magnetic field in finite sized non-ellipsoidal magnetic elements is non-uniform, which introduces additional complexity into the character of the observed magnetization dynamics. For example, the non-uniform demagnetizing field may lead to the spatial confinement and quantization of spin wave modes on the nanometer length scale[9-23]. For thin film elements where the magnetization lies in-plane, the magnitude of the static in-plane demagnetizing field and the non-uniformity of the total effective field acting upon the magnetization increases when the element aspect ratio (size to thickness) is reduced. This results in a richer mode spectrum and hence in a less uniform magnetic response to a pulsed magnetic field, which can be directly imaged in the case of micrometer sized magnetic elements[24-30]. The dominant role of the long range magneto-dipole interaction in the phenomena observed so far makes their analytical description complicated, and generally requires numerical solution of integro-differential equations. However, the interpretation of the magnetization dynamics in non-ellipsoidal elements at finite bias field values becomes even more involved and challenging since it is not only the effective internal magnetic field, but also the static magnetization that is non-uniform. This has led to the development and successful use of numerical algorithms to simulate the spatial dependence and temporal evolution of the magnetization dynamics in thin film magnetic elements[31].

In our previous study of 2.5 nm thick square nanomagnets of less than 200 nm size, we observed a novel dynamical regime in which the response to a uniform pulsed magnetic field was dominated by non-uniform precessional modes localized near the edges perpendicular to the direction of the bias field[15,16]. Micromagnetic simulations of magnetization dynamics in thicker elements revealed the possible existence of a greater number of modes of a more complicated character[19], and also suggested that the exact ground state of the magnetization in the nano-elements might be the key factor in determining their dynamical properties.

In order to explore the effect of the magnetic ground state upon the mode spectrum of nano-elements in more detail, we now present measurements on a set of samples that are similar in composition, shape and lateral size to those in References 15,16, but have about five times greater total thickness, and hence are characterized by a greater non-uniformity of the static magnetization and the total effective field within the element. We find that the precessional mode spectra vary in a discontinuous and complicated fashion as the bias magnetic field is reduced. Good agreement may be obtained between the experimental spectra and those obtained from micromagnetic simulations when the following features are included in the simulations. Firstly, imperfections of element shape are included in the simulations. Secondly, model arrays of 3×3 elements are studied so that the magnetic environment of the center element is correctly modeled. Third, the static magnetization state of all elements in the model array is carefully prepared by applying a similar bias field history to the model arrays as in the experiment. All three factors are found to be important in correctly reproducing the experimental spectra and ultimately understanding the evolution of the spatial character of modes excited in elements of different size and at different values of the bias field.

Despite the agreement between the experimental and simulated spectra, the assignment of the mode character becomes far less clear than in the case of the thinner elements in Reference 16. In particular, we find that the widely adopted concept of the spin wave potential well[11] does not always correctly describe the observed behavior. A general feature of the results presented in this paper is the occurrence of a crossover from a quasi-uniform excitation that occupies the majority of the element volume to a regime that is dominated by several modes weakly localized near the element edges that lie perpendicular to the bias field. The transition is found to be mediated by the increasingly non-uniform static magnetization at low bias field values, and by the associated non-uniformity of the total effective field throughout the element. This new feature of the magnetization dynamics in non-ellipsoidal nano-elements is exhibited for all elements sizes as the bias field is reduced and has not previously been reported. Furthermore, we find that the frequencies of modes belonging to the lower frequency branch are very sensitive to the non-uniform ground state. Indeed, from our measurements, we are able to infer that the elements studied here occupy the S-state. Our experimental spectra also reveal that the lower frequency branch exhibits significant broadening of the linewidth, particularly at low bias fields and/or small element sizes.

The increased thickness, the associated enhanced stray magneto-dipole field, and more non-uniform magnetization are also known to result in a stronger interaction of magnetic elements within arrays[10,18]. The interaction has manifested itself before as extrinsic configurational anisotropy[10] and as correlated switching of individual nano-elements within arrays[32]. It has also been shown to lead to the splitting of normal modes of individual circular elements of 200 nm diameter within an array with a 50 nm edge-to-edge separation[18], which has been interpreted as the formation of collective spin wave modes of the arrays[33]. Our simulations suggest that the increased linewidth of the lower frequency modes observed in our experiments may be associated with the presence of these collective modes.

## 2. METHODS

### 2.1 Experimental technique

For the measurements presented here, we used the same time-resolved scanning Kerr microscope (TRSKM) and methodology as in Reference 16. Square ~ 4×4 μm$^2$ arrays of square elements of length (separation) 637 (25), 428 (17), 236 (77), 124 (30), and 70 (37) nm were studied. The arrays were formed from a Ta(50 Å)/Co$_{80}$Fe$_{20}$(40 Å)/Ni$_{88}$Fe$_{12}$(108 Å)/Ta(100 Å) film sputtered onto a Si wafer and patterned by electron beam lithography and ion milling. Uniaxial anisotropy was induced in the sheet material by field annealing prior to the post-deposition array fabrication. In order to perform pump-probe measurements, two 300 nm thick Au transmission lines with width and separation of 30 μm were fabricated either side of the arrays parallel to the field annealing direction. A GaAs(substrate)/Au(300 nm) photoconductive switch was connected to one end of the transmission lines, while a bias voltage of ~ 20 V was applied to the other end. Optical gating of the photoconductive switch using the pump laser pulse caused a current pulse to propagate along the transmission lines, which in turn generated a pulsed magnetic field around the transmission lines. The pump-probe experiments were performed at a wavelength of 785 nm. The probe beam was linearly polarized and then expanded by a factor of 10 to reduce the beam divergence. A microscope objective of numerical aperture 0.65 was used to focus the probe beam to a diffraction limited spot with a sub-micrometer diameter. A piezoelectric scanning stage was used to position the center of the array under the probe spot. In this configuration, the measured polar Kerr rotation was proportional to the change in the out-of-plane component of the dynamic magnetization. The measured signal was an average response from an ensemble of elements at the center of the array within the probed region. The amplitude of the measured Kerr rotation was typically a few tens of microdegrees. A static bias magnetic field was applied in the plane of the sample parallel to the field annealing direction and the tracks of the stripline structure. The bias field was set by reducing the field from a value of 1.5 kOe that far exceeded the saturation field observed in in-plane hysteresis loops acquired from these arrays using vector Kerr magnetometry[34].

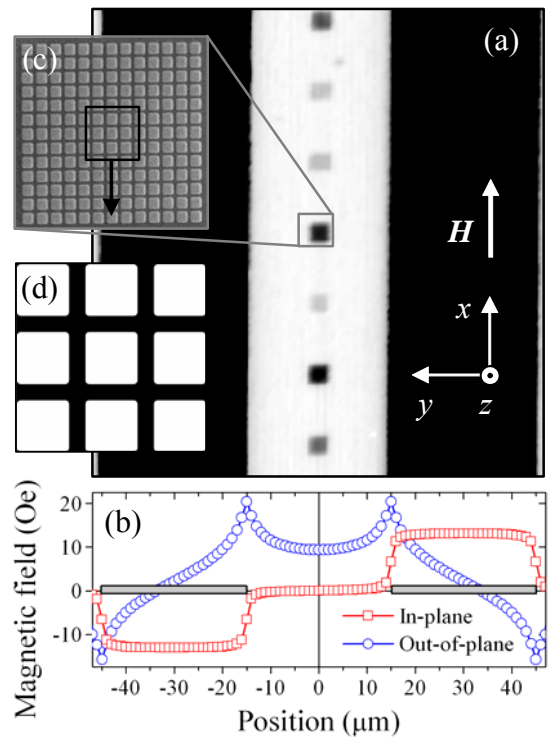

Figure 1 (Color online) (a) An optical micrograph of the sample is shown. (b) The calculated profile of the in-plane (open-squares) and out-of-plane (open-circles) components of the pulsed magnetic field is shown. (c) An SEM image of the array of 236 nm square elements is shown. (d) A 3×3 model array corresponding to the SEM image in (c) and used in micromagnetic simulations is shown. The inset in (a) shows the coordinate system used in the paper.

A continuous film reference sample was co-deposited onto a 1" diameter glass substrate. Vibrating sample magnetometry (VSM) measurements made upon reference samples with different bilayer thicknesses revealed that 12.1 Å of the $Ni_{88}Fe_{12}$ layer was lost due to interdiffusion with the Ta capping layer[16]. From the VSM measurements the static magnetizations of the $Co_{80}Fe_{20}$ and $Ni_{88}Fe_{12}$ were deduced to be 1445 and 585 emu/cm$^3$, respectively. As the layers were strongly exchanged coupled, the bilayer was considered as a single layer with thickness of 13.6 nm, which is equal to the sum of the thicknesses of the constituent layers after accounting for the effect of the interdiffusion. The magnetization of the bilayer was 838 emu/cm$^3$ and was taken to be the volume-weighted average of the individual layer saturation magnetizations. Similarly to Reference 16, time-resolved measurements made upon a co-deposited 10 $\mu$m square allowed the uniaxial (9450±677 erg/cm$^3$) and surface (-0.491±0.018 erg/cm$^2$) anisotropy parameters to be extracted by fitting the data to a macrospin model in the quasi-alignment approximation and assuming a value of 2.1 for the *g*-factor. The easy axis was found to be canted by about 4° from the direction of the striplines. The material parameters deduced here were used in micromagnetic simulations, which are described later. The validity of the assumption of quasi-alignment was verified by performing dynamical macrospin simulations in which the ground state or static orientation of the magnetization was determined using the steepest descent method.

An optical micrograph obtained by scanning the sample beneath the focused laser beam is shown in Figure 1(a). The different grayscale intensity of the different arrays corresponds to different packing fractions. Figure 1(b) shows the calculated profile of the in-plane and out-of-plane components of the pulsed magnetic field[35]. Assuming a current of 64 mA of uniform density in the cross-section of the striplines, the calculation revealed that the out-of-plane pulsed magnetic field was uniform to within 0.4% over the area of the array. Figure 1 (c) shows a scanning electron microscope (SEM) image of the array of 236 nm square elements with a packing fraction of 56%. From the image, one can see that the elements have slightly rounded corners. The SEM images of the studied elements were used to determine the radius of rounding of their corners and the inter-element separations. Figure 1(d) shows a 3×3 model array that corresponds to the SEM image in Figure 1(c) and represents one of those used in the micromagnetic simulations. The inset in Figure 1(a) shows the coordinate system used in the analysis and the direction of the bias field *H*.

## 2.2 Numerical simulations

Micromagnetic simulations have been performed for all samples studied experimentally using the Object Oriented Micromagnetic Framework (OOMMF)[36]. Particular attention has been given to imperfections in the shape, magnetic environment and ground state of the elements. Magnetostatic dipolar interactions with the nearest neighboring elements are more significant in the thicker (13.6 nm) elements than for the previously studied[16] thinner (2.5 nm) elements due to the increased magnetic charge on the element edges perpendicular to the bias field. To account for any modifications to the element ground state and splitting of the excited modes due to interelement interactions, 3×3 model arrays of square elements were generated. Details of element size, edge-to-edge separation, element imperfection, and specifically the radius of rounding of element corners were determined from the SEM images and used to construct the model array used in the simulations. The radius of rounding of element corners was found to be 15 nm for all elements sizes.

For each bias field two stages of numerical simulation were performed. In the first stage the ground state of the magnetization was calculated for which the energy of the magnetic system was minimized. The magnetization was relaxed quickly by setting the damping constant of the simulated material to 0.5. To acquire the static state of the magnetization two methods were investigated. In the first method the static state at each bias field was prepared by allowing the magnetization to relax from the uniform state. In this method the bias field was applied along the *x*- direction. In the second method, for the bias field of 1 kOe only, the static state was also prepared by relaxing the magnetization from the uniform state. However, subsequent static states at bias field values less than 1 kOe were then prepared by relaxing the magnetization from the static state obtained at the previous bias field value. In the second method the bias field was applied 4º from the *x*-direction, parallel to the uniaxial easy axis, so that all elements within the model array would have the same static magnetization state.

It is known that OOMMF favors the flower[37] or X-state due to the assumption of zero temperature[20]. However, the S-state[37] can be stabilized at finite temperatures[20]. We have also found that rounding of the element corners can promote the C-state[37] in some elements. Canting the bias field by 4º in the static simulation stabilized the S-state in the majority of elements. It was expected that if all elements occupied the same static state, then the interpretation of the mode spectra and spatial character would be more readily understood. Hence, the second method was used to simulate the static state for all element sizes. In the experiments, the bias field was not intentionally canted with respect to the *x*-direction, although one should expect some misalignment of a few degrees to occur.

The second stage of the micromagnetic simulations was to calculate the dynamic response of the magnetization of the model arrays to a uniform out-of-plane pulsed magnetic field. The pulsed magnetic field used in the simulations was assumed to have a rise time of 40 ps, a decay time constant of 2 ns, and a magnitude of 15 Oe[16]. The damping parameter and *g*-factor were assumed to have values of 0.01 and 2.1 respectively. For the center element within the model array, the nearest neighbor environment was expected to be a good approximation to that of elements in the interior of the real arrays. Therefore, the average time-resolved response of only the center element was compared with that obtained experimentally.

To determine the spatial character of the excited modes, images of the modes were generated from the simulated time-resolved data as in Reference 29. The magnetization at each time delay was recorded as a vector field map. For each pixel, a time-resolved trace was generated from the out-of-plane component of the magnetization. Fast Fourier transforms (FFT) of the time-resolved trace for each pixel revealed the Fourier components that were excited in a particular pixel. The calculated FFT were then used to reconstruct images of the magnitude and phase of the dynamical magnetization. Thus for each mode previously identified from the FFT spectra obtained from the average response of the center element, a corresponding 'Fourier image' was calculated showing the spatial profile of the magnitude and phase of the dynamical magnetization. For clarity, the images of the FFT magnitude are normalized to 75% of the maximum value of the magnitude in the center element.

## 3. RESULTS AND DISCUSSION

### 3.1 Time-resolved signals

Figure 4(a) shows a typical time-resolved signal measured from the 236 nm element array at a bias field of 270 Oe. The oscillations due to the magnetization precession are superimposed on a slowly varying background due to a transient out-of-plane canting of the magnetization which follows the temporal profile of the pulsed magnetic field. This temporal profile was determined from signals measured at a bias field of ~ 1.5 kOe, at which no significant oscillatory response was observed. To isolate the oscillatory part of the signal, this background was subtracted from the time-resolved signals measured at lower field values[5]. Figure 4(b) shows the FFT spectra calculated from the time-resolved signals in Figure 4(a). The peaks in the spectra calculated from firstly the raw signal, and secondly the raw signal with the background subtracted, agree well for frequencies greater than about 5.5 GHz. The background subtraction extends the range of observable frequencies down to about 1 GHz.

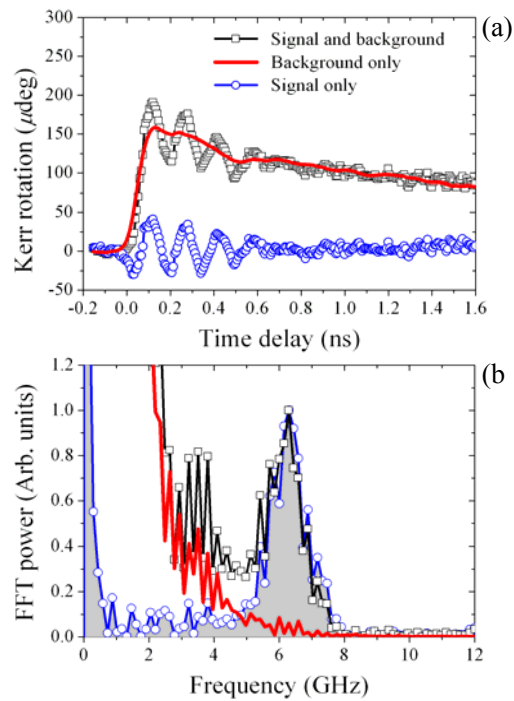

Figure 4 (Color online) (a) A typical raw time-resolved signal (line with squares), the slowly varying background (solid line), and the signal with the background subtracted (line with circles) are shown for the array of 236 nm elements at a bias field of 270 Oe. (b) The FFT power spectra of the traces in (a) are shown.

### 3.2 Size dependence

We consider first the dependence of the mode character upon the element size[16]. FFT spectra calculated from the measured and simulated time-resolved signals are shown in Figure 5 for two values of the bias field (a) 1 kOe, and (b) 150 Oe. At the bias field value of 1 kOe, the dependence of the experimental FFT spectra upon the size of the elements closely resembles that observed in Reference 16. As the element size is reduced from 637 to 236 nm, the frequency of the dominant mode initially increases. At an element size of 124 nm, two modes are clearly seen in both the experimental and simulated data. The 'higher frequency mode' continues the trend of increasing frequency as the element size is reduced, however in the 124 nm element its spectral power is reduced. The 'lower frequency mode' is separated from the higher frequency mode by about 5 GHz in the experimental spectra and has slightly higher spectral power. The frequency of the lower frequency mode then increases as the element size is reduced further. The behavior is qualitatively reproduced by the simulated spectra. The spectra show a crossover in mode intensity for an element size of about 124 nm, while for thinner elements (2.5 nm) in Reference 16 the crossover was observed to occur at the larger element size of about 220 nm. In Reference 16, the crossover was interpreted in terms of the relative size of the demagnetized regions within elements with quasi-uniform static magnetization. The size of the demagnetized regions is expected to increase as the thickness of the elements is increased and the aspect ratio reduced. Therefore, one would expect the crossover to occur at a greater rather than smaller element size for the thicker elements in this study when compared to those of Reference 16. However exactly the opposite trend is observed in Figure 5(a) with the crossover occurring at a smaller element size for the thicker elements. This suggests that the spatial distribution of the demagnetizing field is not the dominant factor that determines the character of the excited modes. The assumption of static uniform magnetization may also be incorrect. Indeed, the magnetization is expected to be less uniform in the thicker elements. Alternatively, the discrepancy may provide evidence for an increased magneto-dipole interaction between elements as the elements become thicker.

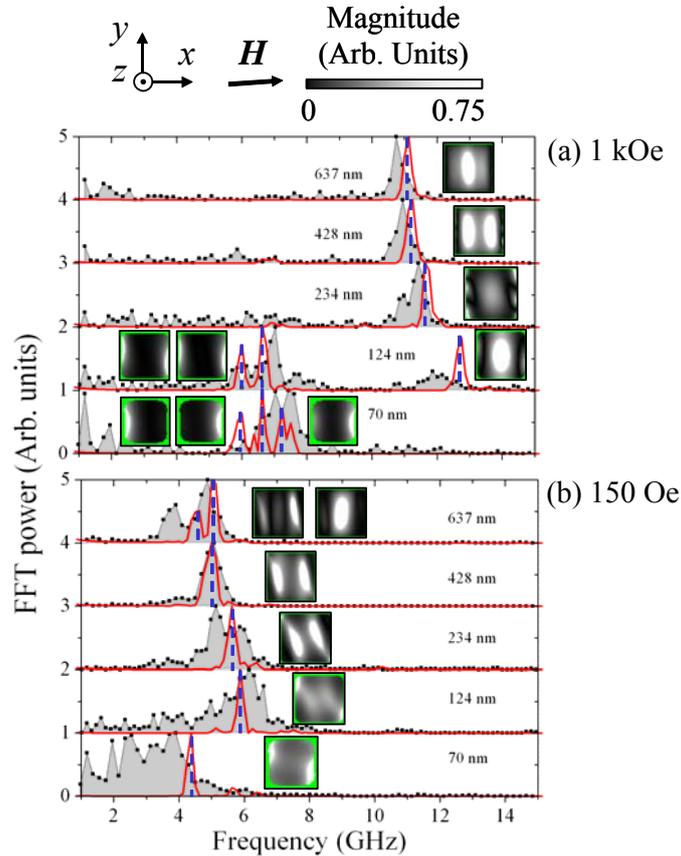

Figure 5 (Color online) The dependence of the mode frequency upon the element size is shown for bias field values of 1 kOe (a) and 150 Oe (b). The shaded spectra were obtained experimentally, while simulated spectra are shown with a solid curve. For each element size, the spatial distribution of the FFT magnitude of the modes in the simulated spectra are shown in the insets.

As in Reference 16, the simulated images of Fourier magnitude reveal that modes from the higher frequency branch occupy the majority of the element volume ('center modes'), while the modes of the lower frequency branch are localized near the edges of the element that are perpendicular to the direction of the bias field ('edge modes'). All modes are non-uniform throughout the elements due to the localized nature of the modes, particularly for the edge modes of the lower frequency branch. Non-uniformity of the center modes from the higher frequency branch occurs in the direction of the bias field, as was previously observed in Reference 19. In Reference 19 a backward-volume type mode was introduced, since the effective wave vector is parallel to the magnetization. The non-uniformity arises because the center mode and backward volume type mode have similar frequencies and finite linewidths and therefore cannot be completely resolved. Also for the three largest elements non-uniformity exists in the direction orthogonal to the bias field at the edges of the element that are perpendicular to the bias field.

At the bias field of 150 Oe, the behavior is remarkably different to that at 1 kOe. Again, the simulations reproduce the experimental data qualitatively, which allows us to use them to discuss the spatial character of the observed modes. The modes do not fall into the classification used in References 16,19 and in the previous paragraph. At 150 Oe, only the 637 nm element supports a center mode with spectral power that is greater than all other excited modes. However, a lower frequency mode is observed with spectral power of about 50% of that of the center mode. In the image of the lower frequency mode the regions of high amplitude appear to be tilted relative to, and detached from the element edges that lie perpendicular to the bias field. We will refer to these modes as 'detached-edge modes'. When the element size is reduced to 428 nm we see that the mode with greatest spectral power has a similar spatial character to the lower frequency mode of the 637 nm element. This is again the case in the 236 nm element. Thus at 150 Oe we observe the size dependent crossover in mode spatial character between element sizes of 637 and 428 nm. Comparison of the size dependent crossover in Figure 5 at 1 kOe and 150 Oe indicates a strong bias field dependence of this general feature of the dynamical behavior of non-ellipsoidal nanomagnets. As the element size decreases we again see an increase in the frequency of the excited modes. However, at 124 nm the spatial character of the dominant mode appears to be somewhat different to that of the modes seen in the larger elements. We interpret the modified spatial profile of the mode to be the result of the two detached edge modes (as seen in the element size of 236 nm) merging at the center of the element. The dominant mode frequency is observed to be reduced by about 2 GHz as the element size is reduced from 124 nm to 70 nm, which might be indicative of a transition of the static magnetization state from an S-state to a leaf-state[37]. Consequently, the high amplitude regions have migrated to the edges of the element perpendicular to the bias field and slightly towards opposite diagonal corners where the effective field may be slightly reduced.

So far the modes shown in the Fourier images of Figure 5 have been classified as center, edge, or detached-edge modes. However, it is not clear from Figure 5 how the mode character evolved as the bias field was reduced from 1 kOe to 150 Oe. In light of the clear field dependence of the mode spatial character, it is necessary to investigate and understand the evolution of the mode spatial character as the bias field is changed. Reducing the bias field for a particular element size allows one to study the effect of the evolution of the static magnetization, total effective magnetic field, and interelement dipolar coupling. The increased non-uniformity of the magnetic ground state in individual elements and/or the increased interaction between elements within the arrays will be shown to be important contributions to the evolution of mode spatial character.

In order to understand the evolution of the spatial character, images of the ground state magnetization have been simulated. At 1 kOe and for all element sizes the magnetization was nearly saturated, but exhibited features of the S-state where the magnetization near the edges of the elements perpendicular to the bias field was slightly canted towards the *y*-direction. Canting of the bias field 4º from the *x*-direction has stabilized the S-state in the center element *and* the surrounding nearest neighboring elements (not shown) for almost all element sizes and bias field values. At 150 Oe, all elements are seen to be in the S-state. The images reveal that a transition from a non-uniform S-state to a similar, but more uniform leaf-state occurs between element sizes of 124 and 70 nm and bias fields of 75 and 0 Oe. Furthermore, for the 3×3 array of 70 nm elements, one of the corner elements acquires the C-state (not shown) for bias fields of 770 Oe and below. Further canting of the bias field to 5º may eliminate this, however one corner element in the C-state has a minimal effect on the behavior of the center element so further simulations were not performed.

As in Reference 16, the linewidth of the experimental spectra for the array of 70 nm elements at a bias field of 1 kOe is noticeably greater than that in the larger elements. The simulations suggest that this may be due to several excited modes with frequencies between 5 and 8 GHz that may not have been resolved in the experiment. In line with this interpretation, the linewidth of the experimental spectra measured at 150 Oe is increased for all element sizes, which is echoed in the simulations by the excitation of multiple 'satellite' peaks around the main spectral peak albeit with rather small spectral power. At the same time, the failure to resolve the modes experimentally should be attributed to the random nature of edge defects and rounding of element corners that give rise to inhomogeneous broadening due to variations within the array[16]. To verify this assumption, we performed micromagnetic simulations for nine isolated elements. Each element had different edge conditions, which included random defects and rounding of corners. The nominal length of all nine elements was 236 nm. A bias field of 590 Oe was applied 4º from the *x*-direction for which the static magnetization of all elements was found to occupy the S-state. The simulations support the interpretation of the inhomogeneous broadening, revealing that slight changes in the static state of the magnetization due to edge defects can cause the mode frequency to vary noticeably even though the shape of the elements is only slightly varied. In some cases the edge defects were found to result in the splitting of modes. Figure 5 suggests that modes localized near to the edges in the elements of smaller sizes appear to be affected more. Two additional simulations were performed for which defects were only included along edges of the element parallel to the bias field (case 1), and along edges of the element perpendicular to the bias field (case 2). The simulated spectra were compared to the spectrum shown in Figure 2(c) for an element with rounded corners and no defects. In all three cases the rounding of the element corners was identical. The spectra revealed that defects along the edges of the element perpendicular to the bias field (case 2) resulted in the largest variation in the mode frequencies. The variation of mode frequency seen in the spectra was most significant for the edge mode. The edge mode was found to split into two modes each of lower frequency than that of the lower frequency mode in the spectrum of Figure 2(c). The simulated spectra presented later in this paper do not exhibit inhomogeneous broadening since the FFT spectra are calculated from the response of only the center element of the 3×3 arrays.

### 3.3    Field Dependence

In light of the strong dependence of the mode structure upon the exact shape of nano-elements, it is difficult to correlate the static and excited states of the magnetization through a comparison of elements of different size and hence slightly different shape. On the other hand, the effective field and the static magnetization state within a particular element may be varied continuously by adjusting the bias field. Since the edge profile is the same for all field values, it is advantageous to concentrate upon the field dependence of the precessional modes.

In Figures 7-11, the dependence of the precessional mode spectra upon the bias magnetic field is shown for elements sizes of 637, 428, 236, 124, and 70 nm, respectively. For the element size of 637 nm (Figure 7) the experimental and simulated spectra show a monotonic decrease of mode frequencies in the range 12 GHz to 2 GHz as the bias field is reduced from 1 kOe to 0 Oe. The experimental and simulated spectra show evidence of lower frequency modes at bias fields of 770 Oe and below, albeit with varying spectral power. The Fourier images in the right-hand panel again show the spatial character of the simulated modes. The vertical blue dashed lines indicate the frequencies of the spectral peaks to which the Fourier images correspond. The images reveal that the mode with the largest spectral power between bias fields of 1 kOe and 150 Oe is a center mode. At 75 Oe the two detached-edge modes possess greater spectral power than the center mode. At 0 Oe the spectral power of the center mode vanishes and the dominant excited modes are two detached-edge modes. At a bias field of 150 Oe, we observe a field-dependent crossover of the spatial character of the modes in the simulations. The experimental spectra do not clearly reveal the crossover, however they do reveal that more than one mode is present. It is also interesting to note that modes with backward-volume type character, with a greater number of nodes and hence a greater wave number, have higher frequency than the quasi-uniform mode. This means that the dominant contribution to their precessional frequency is the exchange interaction.

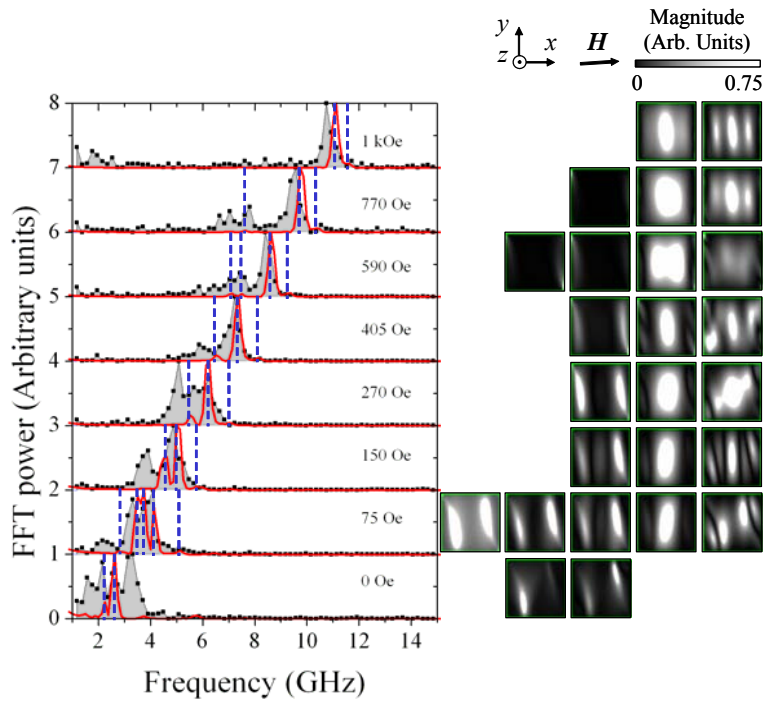

Figure 7    (Color online) The dependence of the mode spectra and the spatial character of the different modes upon the bias field is shown for the 637 nm element. The shaded spectra with symbols show the experimental data, while the solid red line shows spectra obtained from micromagnetic simulations. The vertical dashed blue lines indicate the frequencies of the modes whose spatial character is shown in the images of the magnitude of the Fourier transform. The images are ordered in terms of increasing frequency .

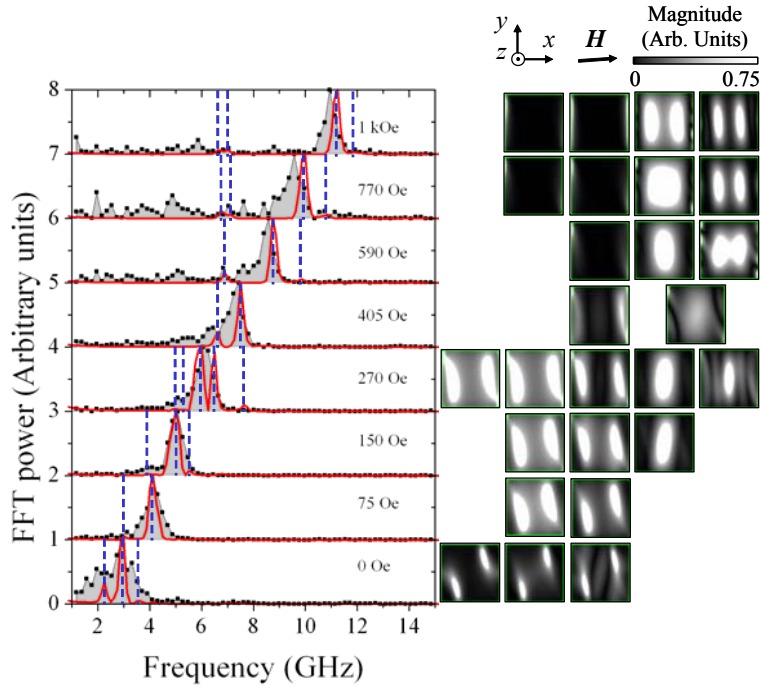

Figure 8  (Color online) The dependence of the mode spectra and the spatial character of the different modes upon the bias field is shown for the 428 nm element. The shaded spectra with symbols show the experimental data, while the solid red line shows spectra obtained from micromagnetic simulations. The vertical dashed blue lines indicate the frequencies of the modes whose spatial character is shown in the images of the magnitude of the Fourier transform. The images are ordered in terms of increasing frequency .

For the element size of 428 nm (Figure 8), the field dependence of the mode character is similar to that of the 637 nm element. Again the experimental and simulated spectra show a monotonic decrease of mode frequencies in the range 12 GHz to 2 GHz as the bias field is reduced from 1 kOe to 0 Oe. At a bias field of 1 kOe, evidence of lower frequency edge modes can be seen in both the experimental and simulated spectra. Also, interference between the high-frequency backward-volume type mode and the center mode leads to a Fourier image that is non-uniform at the center of the element with two anti-nodes. This composite center mode continues to have the largest spectral power until the bias field is reduced to 270 Oe, where a detached-edge mode becomes the mode with largest spectral power. At 150 Oe the spectral power of the center mode is greatly reduced and the detached-edge modes dominate the dynamical response as the bias field is reduced further. From the simulations we see that the field dependent crossover of mode spatial character has taken place at a higher bias field value of 270 Oe.

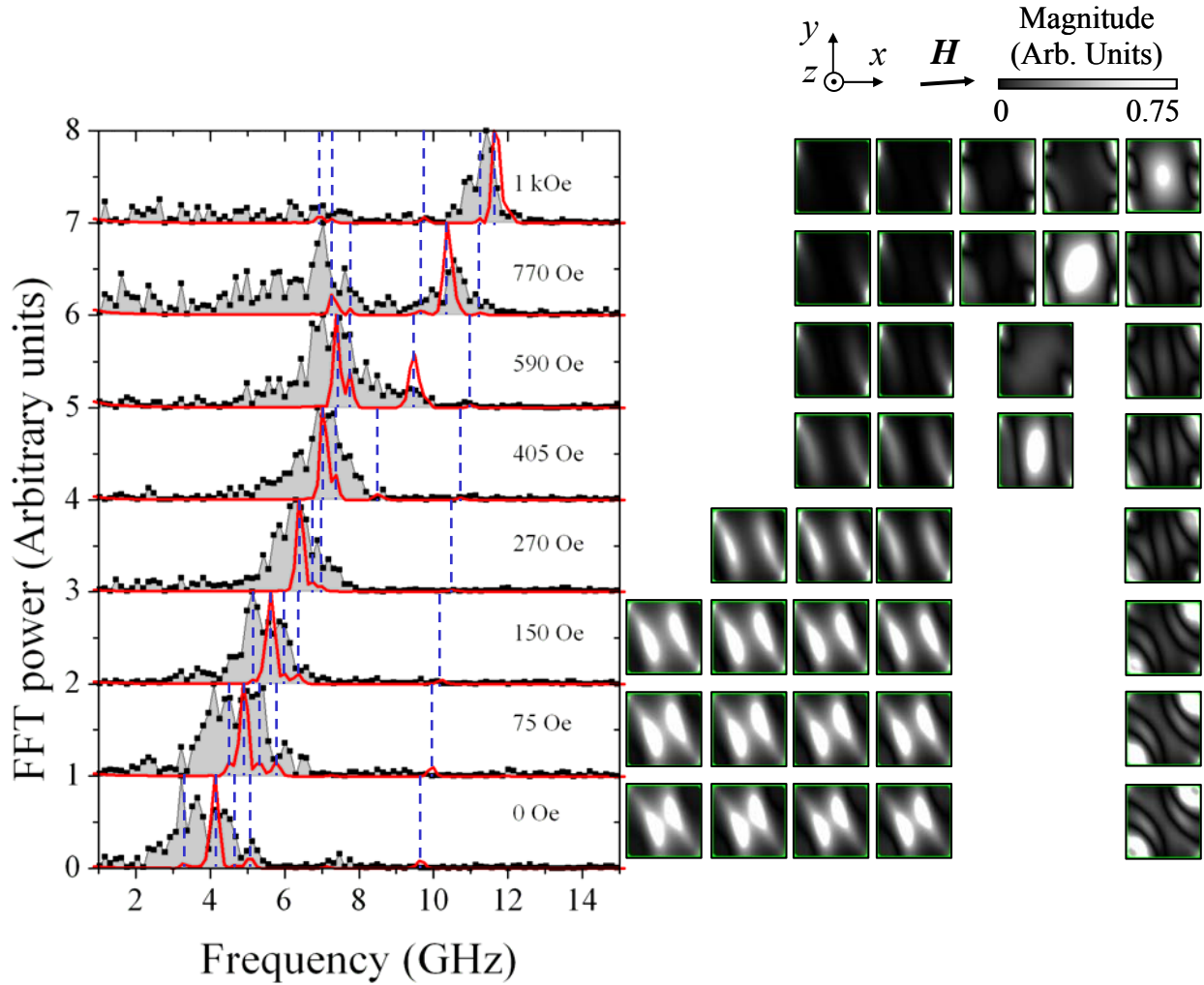

Figure 9 (Color online) The dependence of the mode spectra and the spatial character of the different modes upon the bias field is shown for the 236 nm element. The shaded spectra with symbols show the experimental data, while the solid red line shows spectra obtained from micromagnetic simulations. The vertical dashed blue lines indicate the frequencies of the modes whose spatial character is shown in the images of the magnitude of the Fourier transform. The images are ordered in terms of increasing frequency .

For the element size of 236 nm (Figure 9), a crossover from a center mode to a detached-edge mode occurs between bias fields of 770 and 590 Oe. The experimental spectra show that the lower frequency edge mode has greater spectral power at 770 Oe, while the simulated spectra suggest that the edge mode becomes dominant at 590 Oe. As the bias field is reduced from 1 kOe to 590 Oe, the lower frequency edge modes do not exhibit a monotonic decrease in frequency, but instead increase in frequency by about 0.5 GHz. As for the 637 and 428 nm elements, the frequency of the center mode of the 236 nm element exhibits a monotonic decrease as the bias field is reduced, before the mode amplitude is greatly attenuated at 270 Oe. Also, between bias field values of 1 kOe and 405 Oe the Fourier images reveal that the center mode is non-uniform. At all bias field values the simulated spectra and Fourier images reveal a non-uniform mode between 9.6 and 10.3 GHz. Similarly for bias field values of 1 kOe to 590 Oe a non-uniform mode exists between 9.8 and 9.5 GHz. Between 1 kOe and 590 Oe at least one of the non-uniform modes has a similar frequency to the center mode. As seen for the 637 and 428 nm elements it is not possible to completely resolve modes with similar frequencies since the spectral lines have a finite linewidth. At 590 Oe the mode with frequency of about 9.8 GHz at 770 Oe seems to merge with the center mode spectral peak. The corresponding Fourier image reveals significant non-uniformity of the high amplitude region at the center of the element. At 590 Oe the experimental spectral peak is very broad and is accompanied by two modes of similar frequency in the simulated spectrum. The inhomogeneous broadening is again attributed to the random nature of defects and irregularities at the edges of the elements, which causes dispersion of the frequency of the excited modes near to element edges due to variations of the local effective field and pinning of the magnetization. The Fourier images corresponding to simulated spectra for bias fields of 590 Oe and below reveal that the modes of different frequency are in fact weakly associated with the element edges. Finally, the simulated spectra reveal that for all bias field values, lower frequency modes with different frequency possess very similar, if not the same, spatial character, which is the result of collective-type excitations across the 3×3 array (discussed later).

For the element size of 124 nm (Figure 10), a crossover from a center mode to an edge mode occurs at 1 kOe. Between bias fields of 770 and 270 Oe the lower frequency (< 8 GHz) experimental spectra are very broad. The simulated spectra show that there are many lower frequency modes (up to 6) excited with variable frequency and spectral power. In line with the previous discussion, smaller elements will be more susceptible to frequency dispersion and inhomogeneous broadening of spectra since the effects of edge defects and irregularities are more pronounced. Furthermore, the excited modes in smaller elements seem to be mostly edge-type modes, which again are more susceptible to edge conditions. Between 1 kOe and 590 Oe there seem to be two center modes with different frequency, which are the result of collective-type excitations (discussed later). Similarly, many of the lower frequency edge-type modes share the same spatial character, but again with different frequency. Between 270 Oe and 75 Oe the high amplitude regions of the edge-type modes have extended into the center of the element, while at 0 Oe the high amplitude region has become detached from the element edges.

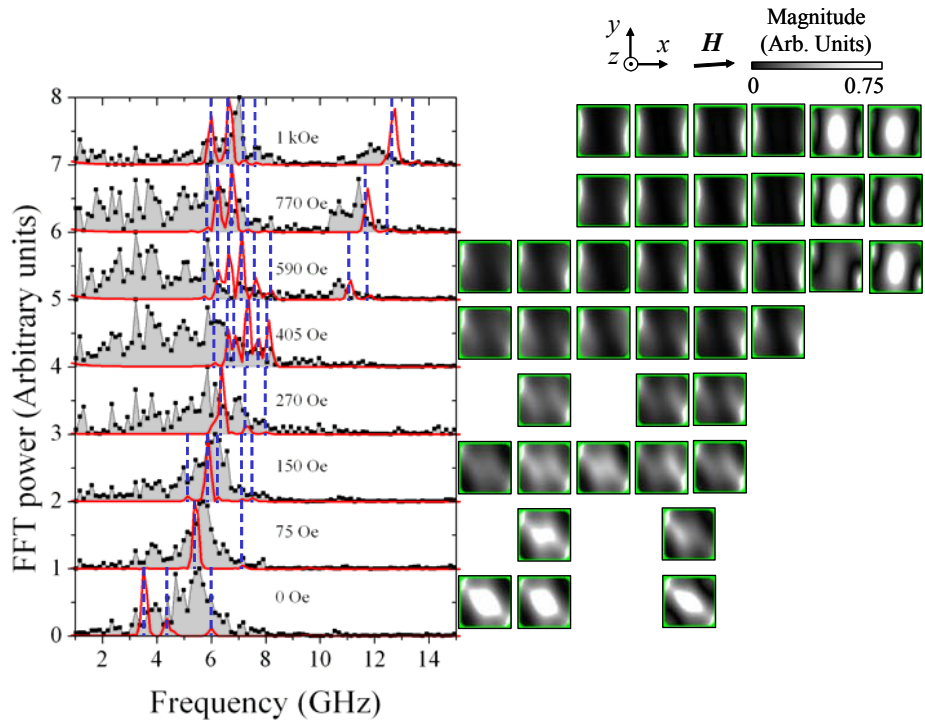

Figure 10   (Color online)  The dependence of the mode spectra and the spatial character of the different modes upon the bias field is shown for the 124 nm element. The shaded spectra with symbols show the experimental data, while the solid red line shows spectra obtained from micromagnetic simulations.  The vertical dashed blue lines indicate the frequencies of the modes whose spatial character is shown in the images of the magnitude of the Fourier transform.  The images are ordered in terms of increasing frequency .

This may be interpreted as being due to a change of the static magnetization configuration.  In the simulated spectra a large change in frequency (~ 2 GHz) is seen between 75 Oe and 0 Oe which is not so apparent in the experimental spectra.  Indeed simulated images of the static magnetization (Figure 6) and the effective field (shown and discussed later) reveal a transition from an S-state to a leaf-state.

For the element size of 70 nm (Figure 11), a crossover of mode character is not observed when the bias field is reduced.  The crossover may be revealed in micromagnetic simulations performed at a bias field value above 1 kOe.  The experimental and simulated spectra show that only lower frequency (< 8 GHz) modes are excited for bias fields of 1 kOe and below.  That is, the higher frequency branch of the center mode is not observed in either the experimental spectra or simulated spectra. The corresponding Fourier images reveal that all modes excited between 1 kOe and 150 Oe are edge-type modes. Again there is significant inhomogeneous broadening of the experimental spectra, which is supported by the presence of many modes with slightly different frequency in the simulated spectra. As seen for the 124 nm element, the high amplitude regions of the 70 nm element have extended into the center of the element between 75 Oe and 0 Oe. Again the simulated static magnetization and effective field reveal a transition from an S-state to a leaf-state between these bias field values.

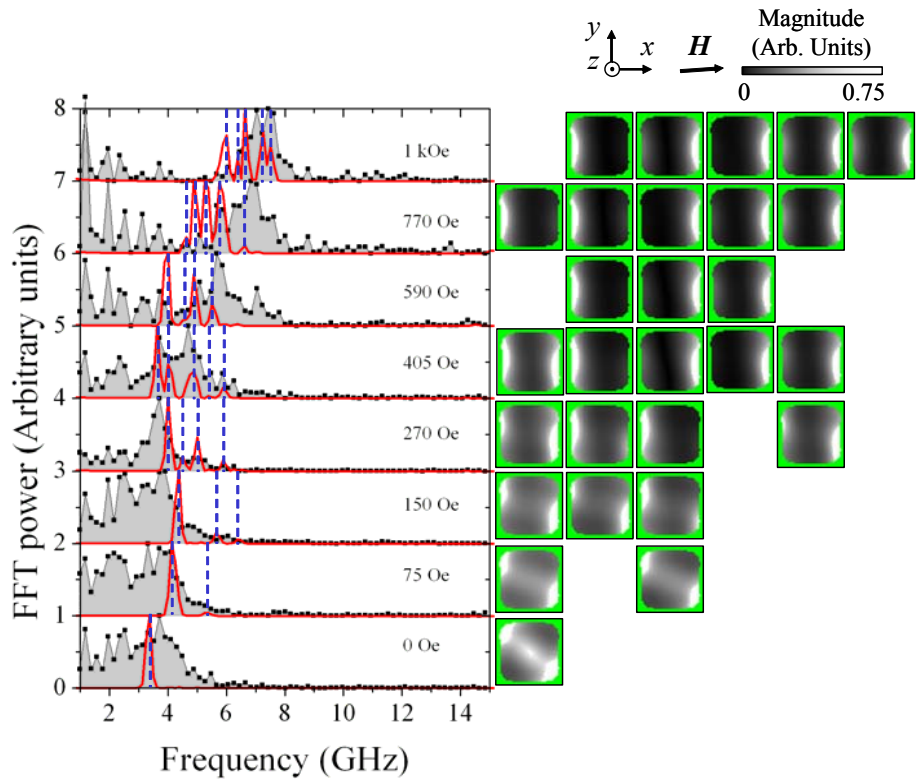

Figure 11    (Color online) The dependence of the mode spectra and the spatial character of the different modes upon the bias field is shown for the 70 nm element. The shaded spectra with symbols show the experimental data, while the solid red line shows spectra obtained from micromagnetic simulations. The vertical dashed blue lines indicate the frequencies of the modes whose spatial character is shown in the images of the magnitude of the Fourier transform. The images are ordered in terms of increasing frequency.

The static state of the magnetization becomes increasingly non-uniform as the bias field is decreased. In Figure 13 images of the total effective field ($H_{eff}$) corresponding to the magnetization configurations in Figure 6 are shown. The effective field includes contributions from the applied, demagnetizing, anisotropy, and exchange fields. The grayscale represents the magnitude of the effective field normalized to the bias field. The images clearly show that a reduction of either the element size or the bias field can result in a non-uniform effective field. At 1 kOe, the effective field is nearly uniform at the center of all elements (light gray). For the 637 and 428 nm elements, regions of small internal field exist in narrow regions along the edges perpendicular to the applied field (black). As the element size is reduced from 236 to 70 nm the regions of small internal field move towards the center of the element and remain perpendicular to the bias field. Similarly, when the bias field is reduced from 1 kOe the regions of small internal field within the 637 and 428 nm elements begin to move towards the center of the element at 405 Oe. However, at bias field values of 405 Oe and below the effective field becomes non-uniform in both *x*- and *y*-directions. At 75 Oe regions of negative internal field can clearly be seen from the direction of the arrows. A similar evolution of the effective field is seen for the 236 and 123 nm elements as the bias field is reduced, however it seems that regions of small internal field are observed at bias field values of 1 kOe and greater. At a bias field value (element size) of 0 Oe (637 and 428 nm), 150 Oe (236 nm), 270 Oe (123 nm), and 590 Oe (70 nm), the effective field becomes negative throughout the majority of the element. The most significant contribution to the effective field is then the demagnetizing field and the regions of small internal field (black) are expelled from the element near the edges parallel to the bias field.

For the 637, 236, and 70 nm elements respectively, Figures 15, 16, and 17 show cross-sections of (a) the effective field images (shown in Figure 13) and (b) the FFT magnitude of the dominant edge and center modes, where both appear, (shown in Figures 7, 9, and 11) for four different bias field values. Generally there are three regions of interest in the cross-section of the effective field, i) a region of zero or negative effective field adjacent to the edges of the element perpendicular to the bias field (edge region), ii) a region of positive effective field at the center of the element (center region), iii) a region that connects the first two regions where the gradient of the effective field is large (gradient region).

For the 637 nm element, Figure 15(a) reveals that there is a small increase in the width $\Delta x$ of the edge regions of about 2% of the element length as the bias field is reduced from 1 kOe to 270 Oe. At the same time, the magnitude of the effective field in the edge regions decreases from about –750 Oe to –250 Oe. At 75 Oe it can be seen that the width of the edge regions has increased significantly to about 28% of the element length, and a maximum in the negative effective field has formed. Within these edge regions at 75 Oe, the magnitude of the negative effective field is larger than the positive effective field in the center region. As the bias field is reduced from 1 kOe to 75 Oe the effective field in the center region remains approximately equal to the bias field. Furthermore, the width of the gradient regions increases at the expense of that of the center region.

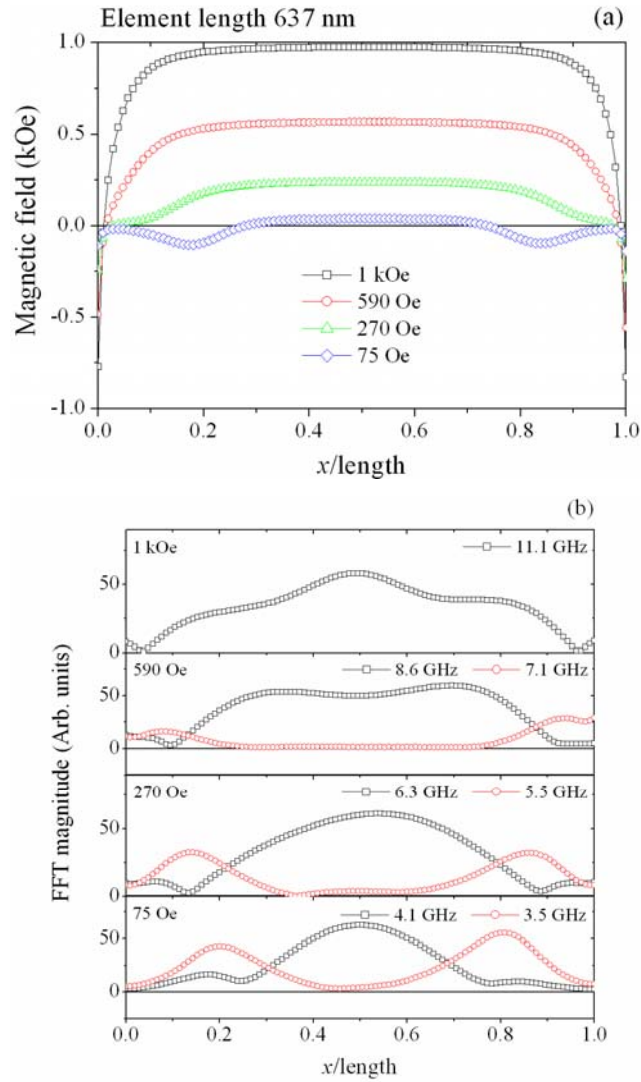

Figure 15    (Color online) Cross-sections are shown of the *x*-component of the simulated total effective field (a) and the FFT magnitude (b) within the center element of the 3×3 model array of 637 nm elements. The sections are shown for four different values of the bias field.

In Figure 15(b) two types of modes are seen from the cross-section of the FFT magnitude images; i) a mode with large FFT magnitude at the center of the element (open black squares), ii) a mode with large FFT magnitude near the edges of the element perpendicular to the bias field (open red circles). It would be incorrect to characterize these mode in terms of strict localization to the center, edge or gradient regions. For example, for a mode to be center (or edge) localized one would expect the FFT magnitude to be equal to zero at the edges (or center) of the element. In Figure 15(b) at 1 kOe the 11.1 GHz mode has two minima in the profile of the FFT magnitude near to the element edges.

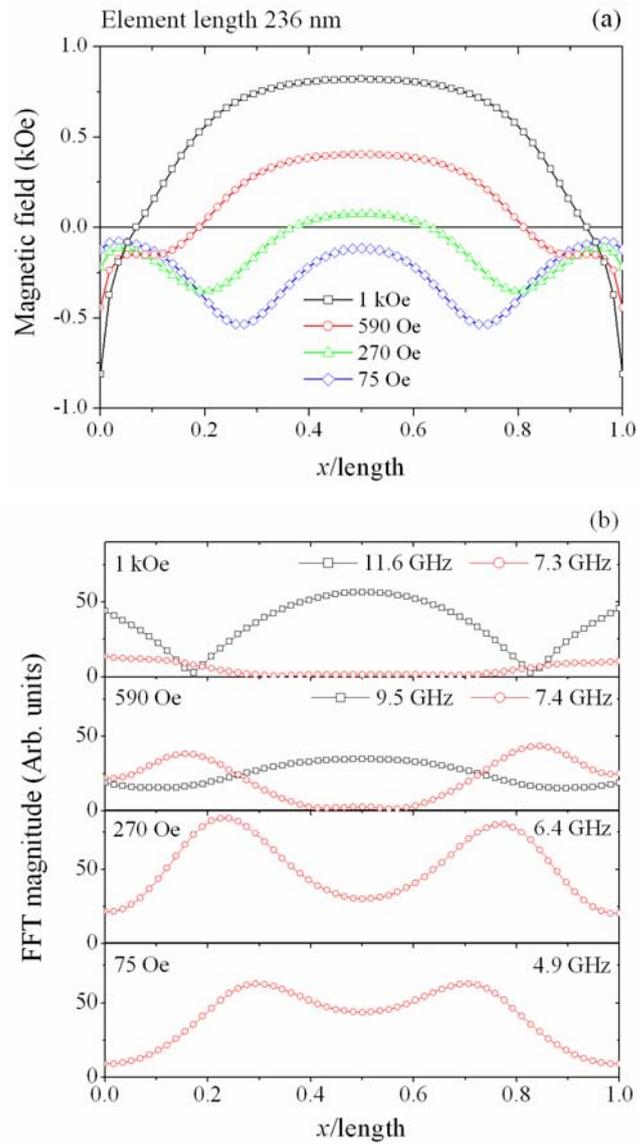

Figure 16 (Color online) Cross-sections are shown of the *x*-component of the simulated total effective field (a) and the FFT magnitude (b) within the center element of the 3×3 model array of 236 nm elements. The sections are shown for four different values of the bias field.

As the bias field is reduced, the minima migrate from positions near the edge of the element at 1 kOe towards the center of the element. At 590 Oe the lower frequency mode (7.1 GHz) has two maxima in the profile of the FFT magnitude that correspond to the position of the two minima of the higher frequency (8.6 GHz) mode. As the bias field is reduced these maxima also migrate towards the center of the element.

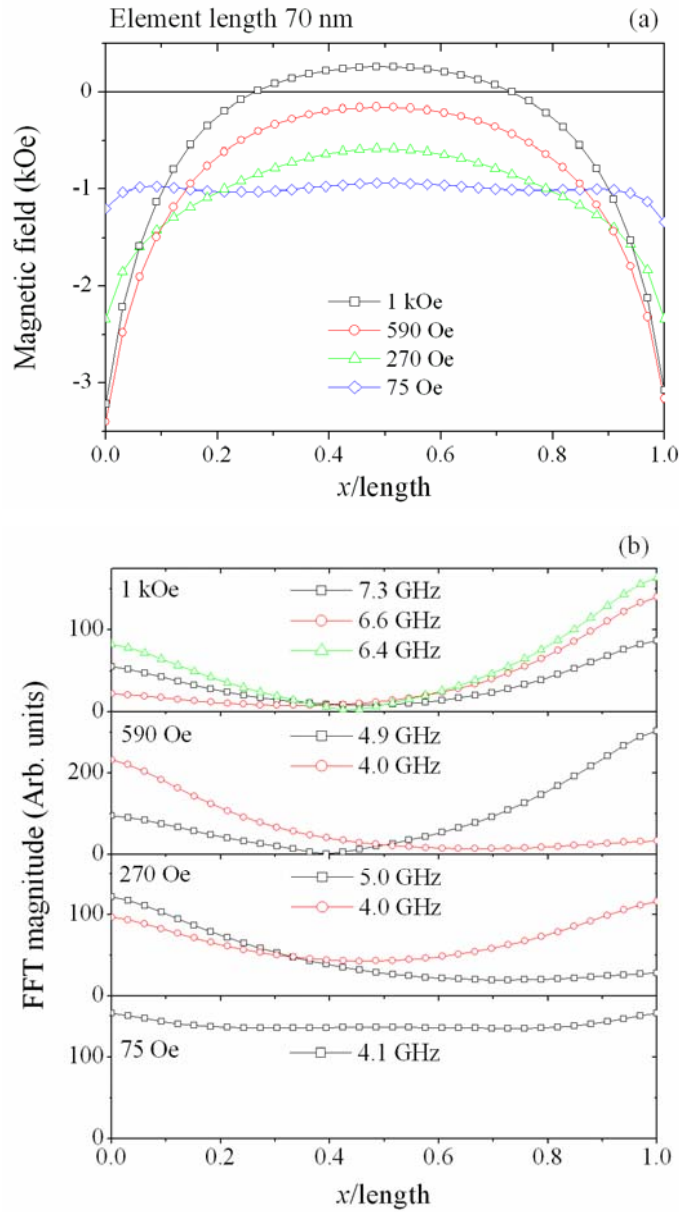

Figure 17   (Color online) Cross-sections are shown of the *x*-component of the simulated total effective field (a) and the FFT magnitude (b) within the center element of the 3×3 model array of 70 nm elements. The sections are shown for four different values of the bias field.

Features in the cross-section of the FFT magnitude for both modes are correlated with those of the effective field in Figure 15(a). In Reference 11 for a transversely magnetized microscale stripe, a gradient region near to the stripe edge was identified as the region of localization for low frequency modes with edge-type spatial character. The edge region (zero field and non-uniform magnetization) was assumed to reflect spin-waves propagating from the center of the stripe towards these regions. Furthermore, a second turning point in the effective field was identified, above which only modes with an imaginary effective wavevector could exist. The region between these two boundaries was identified as the region of localization of the lowest frequency mode. Here, in smaller sub-micron non-ellipsoidal elements the lower frequency modes have large FFT magnitude in the gradient regions, but are not localized there. The FFT magnitude extends across regions of positive and negative internal field, as seen in Figure 15(b) at 590 Oe. The region of large FFT magnitude associated with the higher frequency mode decreases in width as the bias field is decreased. At the same time, the FFT magnitude of the lower frequency mode increases near the edge of the elements and migrates towards the center of the element. However, while the effective field is positive in the center region, the FFT amplitude of the lower frequency mode in that region remains small.

For the 236 nm element at 590 Oe, the effective field (Figure 16(a)) in the center region is positive and the FFT magnitude (Figure 16(b)) of the lower frequency mode is small. However, at 75 Oe, the effective field is negative across the whole section of the element. The regions of high FFT magnitude of the lower frequency mode that were near the element edges now also extend across the whole element. It is clear from Figure 16(a) that the effective field within the center region of the element is no longer similar to the bias field. Instead, the demagnetizing field contributes significantly to the effective field.

For the 70 nm element, the effective field (Figure 17(a)) is only positive in the center region when the bias field is 1 kOe, and is significantly less than the bias field value. As discussed earlier, modes from the higher frequency branch that were found to occupy the center region of larger elements (Figures 7-10) were not observed in the simulation of 70 nm elements at 1 kOe (Figure 11). Instead, at 1 kOe the modes excited within the 70 nm element have large FFT magnitude in the edge regions (Figure 17(b)) where the effective field is large (~ 3 kOe) and negative. In the center region where the effective field is positive, the FFT magnitude of the modes is small. Between the bias field values of 590 and 75 Oe the large demagnetizing field results in a negative effective field across the whole section of the element. Figure 11 reveals a number of different non-uniform modes that that tend to have maximum magnitude at the edges of the element. Figure 17(b) reveals that the FFT magnitude of these modes is non zero throughout the element, with regions of large FFT magnitude correlated with regions of large negative internal field. Finally, at 75 Oe, Figure 6 and 13 reveal that the 70 nm element no longer occupies the S-state, but now occupies the leaf-state. In this case the negative effective field and the FFT magnitude are almost constant across the whole section of the element.

It is clear from Figure 16(b) that the crossover observed in the spectra and Fourier images of Figure 9 is the result of a change in the spatial character of the excited modes. The crossover is mediated by the changes in the effective field throughout the element as the bias field is decreased. However, since the modes have non-zero FFT magnitude throughout, the crossover cannot be simply described as a crossover from a center- to an edge-localized mode character. In a similar manner the change in the effective field that occurs as the element size is reduced (Figure 13) can be used to explain the size dependent crossover of the spatial character of the mode seen in Figure 5.

### 4.2   Collective modes within arrays

In Reference 18 the effect of interelement interactions upon high frequency normal modes was modelled in 3×3 square arrays of circular permalloy dots of thickness 50 nm, diameter 200 nm and interdot separation ranging from 50 to 800 nm. For an interelement separation greater than about 200 nm the high frequency response was found to be successfully modeled by that of an individual element. For interelement separations less than 200 nm, the magnetostatic interaction was found to significantly modify the mode character in the dots. Here, the interelement separation is less than 200 nm for all element sizes, and the separation is not varied. However, by varying the bias field the magnetostatic interaction between elements can be changed and investigated.

So far in this paper we have interpreted the experimental spectra by comparing them with the corresponding simulated spectra for the center element of 3×3 model arrays and then calculating the Fourier images of the spatial character for particular modes. For all element sizes, particularly for the lower frequency modes (< 8 GHz) the spatial character of modes excited in the center element are similar if not the same as other excited modes for a particular element size and bias field. We interpret this as evidence for collective modes within the 3×3 array. The simulated Fourier images of the center element reveal that the spatial character of the FFT magnitude *and* phase were similar for modes with different frequency. However, in these cases the spatial character of amplitude and phase across the 3×3 array were found to be different.

In Figure 18(a), the FFT magnitude and phase are shown for two modes with frequencies of 6.75 and 7.0 GHz excited in the center element of a 3×3 array of 236 nm elements. While there is a finite phase difference between the two modes, the variation of the phase within each image is similar. In Figure 18(b) the FFT magnitude is shown for all elements in the 3×3 array. While the spatial character of the two modes are similar in the center element, the character of the modes across the array are very different demonstrating that collective modes occur within the array. The excitation of collective modes with similar frequencies may contribute to linewidth broadening of the experimental spectra in Figures 7-11. The Fourier images in Figures 7-11 reveal that below the crossover field there are often several modes at a particular bias field with similar spatial character. Indeed, as the bias field is reduced the static state of the magnetization becomes increasingly non-uniform and the magnetostatic interaction between elements increases[10]. Therefore, below the crossover field a larger splitting of the frequencies of the collective modes is expected. This is supported by the increased linewidth observed experimentally at lower bias field values.

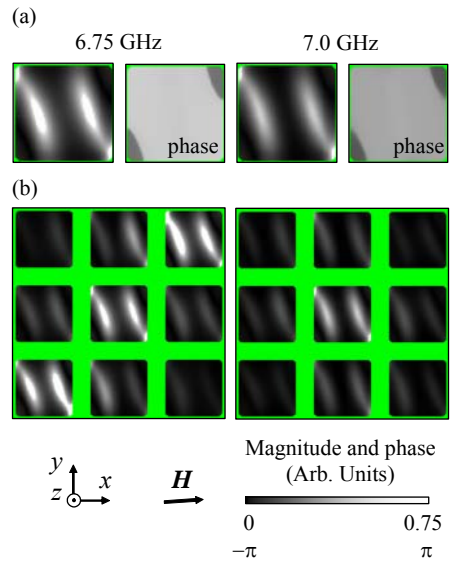

Figure 18  (Color online) Simulated Fourier images of FFT magnitude and phase are shown in (a) for two modes with frequencies of 6.75 and 7.0 GHz excited in the center element of a 3×3 array of 236 nm elements at a bias field of 270 Oe. In (b) the FFT magnitude is shown for all elements in the 3×3 array.

## 5.  SUMMARY

We have performed TRSKM measurements upon arrays of square ferromagnetic nano-elements of different size and for a range of bias fields. The experimental results were compared to micromagnetic simulations of model arrays in order to understand non-uniform precessional dynamics within the elements. Experimentally, two branches of excited modes were observed to co-exist above a particular bias field. Below the so called crossover field, the higher frequency branch was observed to vanish.

Micromagnetic simulations and Fourier imaging revealed the spatial character of the two mode branches. The modes of the higher frequency branch were found to have high FFT amplitude at the center of the element in regions of positive effective field, while modes of the lower frequency branch were found to have high FFT amplitude near the edges of the element perpendicular to the bias field. Cross-sections of the simulated images of the effective field and FFT magnitude revealed that the crossover between the higher and lower frequency branches was mediated by the complicated evolution of the total effective field within the element. Below the crossover field the increase in the width of the edge region, of negative effective field, at the expense of the center region, of positive effective field, allowed the edge-type mode to extend over the entire element. The simulations revealed that the majority of the modes were de-localized with non-zero FFT magnitude throughout the element. Therefore, the spin-wave well model introduced in Reference 11 for micron sized non-ellipsoidal elements could not be used here to characterize the excited modes as strictly center- or edge-localized modes. The de-localized nature of the excited modes seems to be an intrinsic property of sub-micrometer and nanoscale non-ellipsoidal elements. However, the mode spatial character was found to be correlated with features of the effective field and the static magnetization state.

The simulations revealed that the frequency of modes from the lower frequency branch were very sensitive to the static state magnetization of all the elements within the model array. Therefore, by matching the simulated spectra to the experimental spectra we are able to infer that the elements studied here occupy the S-state. Since, below the crossover field the static magnetization state was shown to be non-uniform, increased magnetostatic interaction is expected between the elements within the array. The simulated spectra revealed that many modes may be excited that have similar spatial character within the center element. However, inspection of the entire array revealed the existence of collective modes, where the FFT amplitude in the elements surrounding the center element was considerably different for two such modes. The excitation of collective modes with similar spatial character in the center element but different frequency may account for the increased linewidth observed in the experimental spectra below the crossover field.

Finally, the results presented in this experimental paper may be useful for the development of a thorough analytical theory of non-uniform modes within magnetic nano-elements with non-uniform static magnetization and total effective field. Detailed knowledge of the mode character of square nano-elements and their static magnetization state is also essential for data storage applications in the finite field regime.

## 6. ACKNOWLEDGEMENTS

Financial support for this work was provided by the UK Engineering and Physical Sciences Research Council (EPSRC) and the New Energy and Industrial Technology Development Organization (NEDO).